\documentclass{desyproc}
\usepackage{graphicx}
\begin{document}
\title{SMASH-ing Vacuum Metastability}

\author{{\slshape C.R. Das$^1$, Katri Huitu$^2$ and Timo J. K\"arkk\"ainen$^2$}\\[1ex]
$^1$Bogoliubov Laboratory of Theoretical Physics, Joint Institute of Nuclear Research,\\
Joliot-Curie 6, 141980 Dubna, Moscow region, Russian Federation
\\
$^2$Department of Physics and Helsinki Institute of Physics,\\
P. O. Box 64, FI-00014 University of Helsinki, Finland}

\contribID{Das\_chittaranjan}

\confID{20012} 
\desyproc{DESY-PROC-2018-03}
\acronym{Patras 2018} 
\doi 

\maketitle

\begin{abstract}
Five fundamental problems - neutrino oscillations, baryogenesis, dark matter, inflation, strong CP problem - are solved at one stroke in ``SM-A-S-H" (Standard Model-Axion-Seesaw-Higgs portal inflation) model by Andreas Ringwald et. al. The Standard Model (SM) particle content was extended by three right-handed SM-singlet neutrinos $N_i$, a vector-like color triplet quark $Q$, a complex SM-singlet scalar field $\sigma$ that stabilises the Higgs potential, all of them being charged under a global lepton number (hyper-charge) and Peccei-Quinn (PQ) $U(1)$ symmetry. We found numerically that SMASH model not only solves five fundamental problems but also the sixth problem ``Vacuum Metastability" through the extended scalar sector.
\end{abstract}

\section{Introduction}
It is well-known that the Standard Model (SM) Higgs potential is metastable \cite{1}, as the sign of the quartic coupling $\lambda_H$ turns negative at instability scale around $\Lambda_{IS}\sim 10^{11}$ GeV. The largest uncertainties of SM vacuum stability are driven by both the top quark pole mass and the mass of SM Higgs boson. Experimental current data is in significant tension with the stability hypothesis, making it more likely that the universe is in a metastable vacuum state. The expected lifetime of vacuum decay to a true vacuum is extraordinarily long, and it is unlikely to affect the evolution of the universe. However, it is unclear why the vacuum state entered to metastable or unstable vacuum, to begin with during the early universe.

It is possible that at or below the instability scale heavy degrees of freedom originating from a theory beyond the SM start to alter the running of the SM parameters of renormalization group equations (RGE). This approach aims to solve the vacuum metastability problem by proving that the universe is currently in a stable vacuum. One theory candidate is a complex singlet $\sigma$ extended SM. The scalar sector of such a theory may stabilise the theory with a threshold mechanism \cite{4,5}. The effective SM Higgs coupling gains a positive correction $\delta\equiv \lambda^2_{H\sigma} /\lambda\sigma$ at $m_\sigma$, where $\lambda_{H\sigma}$ is the Higgs doublet-singlet portal coupling and $\lambda_\sigma$ is the quartic coupling of $\sigma$.

This threshold mechanism is embedded in a recent SMASH \cite{2,3} theory, which utilizes it at $\lambda_{H\sigma}\sim -10^{-6}$ and $\lambda_\sigma\sim 10^{-10}$, where the vacuum expectation value $v_\sigma\sim10^{11}$ GeV breaks the lepton number and the Peccei-Quinn symmetry simultaneously. The SMASH framework \cite{2,3} expands the scalar sector of the SM by introducing a complex singlet field $\sigma$. 

\section{Threshold correction}
Consider an energy scale below $m_\sigma<\Lambda_{IS}$, where the heavy scalar $\sigma$ is integrated out. The low-energy Higgs potential should match the SM Higgs potential:
\begin{equation}
V(H) = \lambda^{SM}_H\left(H^\dagger H-\frac{v^2}{2}\right)^2.
\end{equation}
It turns out that the quartic coupling which we measure has an additional term:
\begin{equation}
\lambda^{SM}_H=\lambda_H-\frac{\lambda^2_{H\sigma}}{\lambda_\sigma}.
\end{equation}
Since the SM quartic coupling will be approximately $-0.08$ at $M_P$, the threshold correction
\begin{equation}
\delta\equiv\frac{\lambda^2_{H\sigma}}{\lambda_\sigma},
\end{equation}
should be large enough to push the high-energy counterpart $\lambda_H$ to positive value all the way up to $M_P$. In the literature there are two possible ways of implementing this threshold mechanism.

One may start by solving the SM RGE's up to $m_\sigma$, from where the SMASH effect kicks in, and the quartic coupling $\lambda_H$ gains a sudden increment by $\delta$. Continuation of RGE analysis then requires utilizing SMASH RGE's up to the Planck scale, $M_P=1.22\times 10^{19}$ GeV.

\begin{figure}[t]
\centerline{\includegraphics[width=1.095\textwidth]{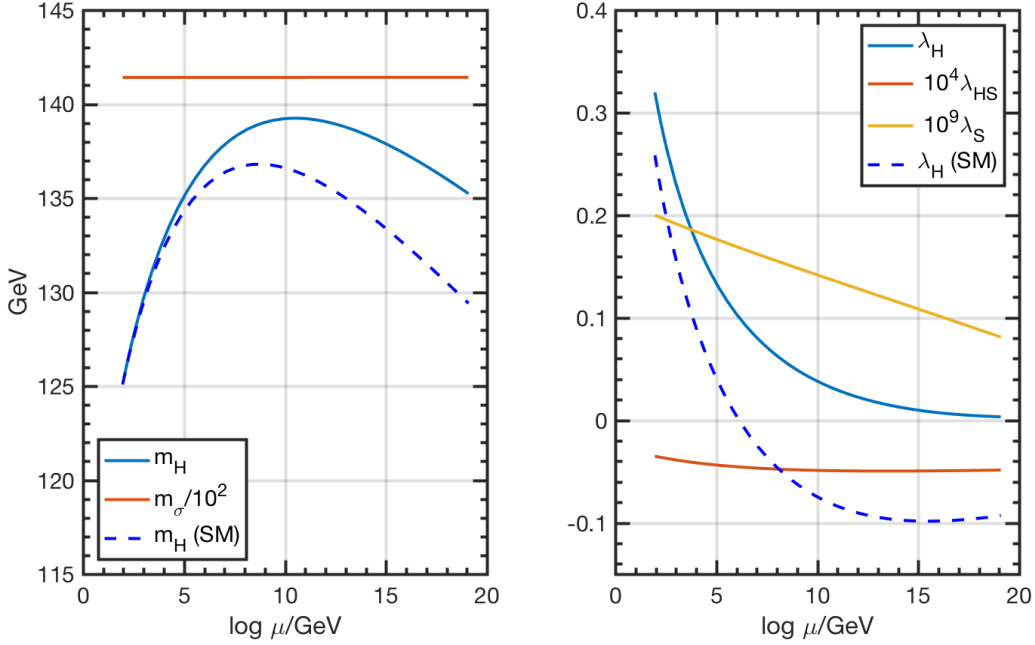}}
\caption{Running of Higgs, $\sigma$ bare mass and scalar potential parameters with benchmark point. Threshold applied from the beginning at $m_Z$.}\label{Fig:1}
\end{figure}

\begin{figure}[t]
\centerline{\includegraphics[width=1.115\textwidth]{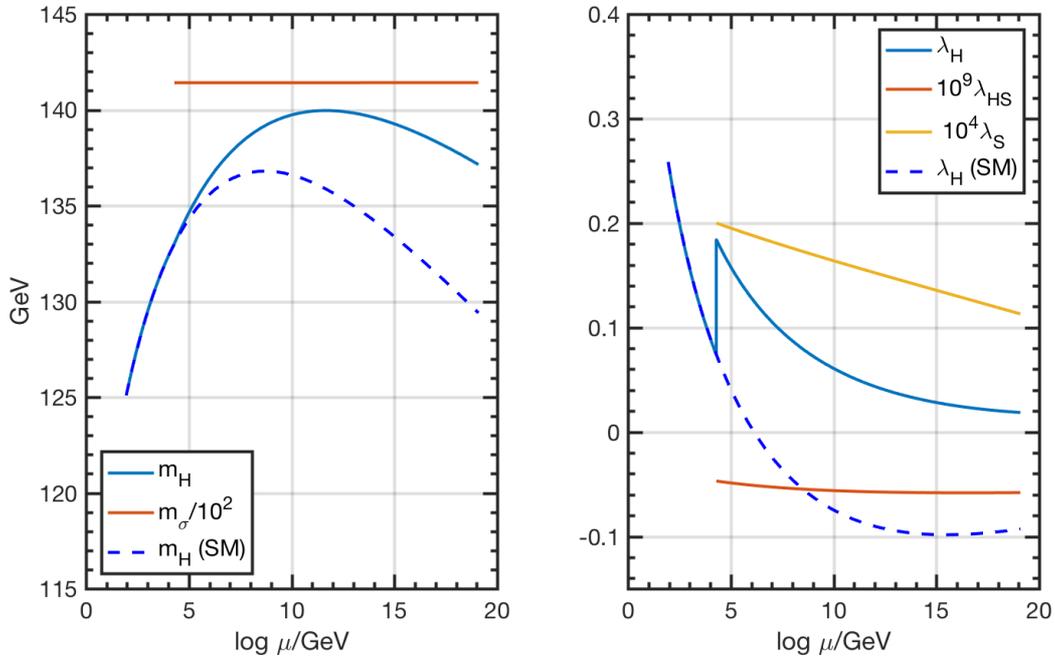}}
\caption{Running of Higgs, $\sigma$ bare mass and scalar potential parameters with benchmark point. Threshold correction utilized at $m_\rho$.}\label{Fig:2}
\end{figure}

\begin{figure}[t]
\centerline{\includegraphics[width=0.515\textwidth]{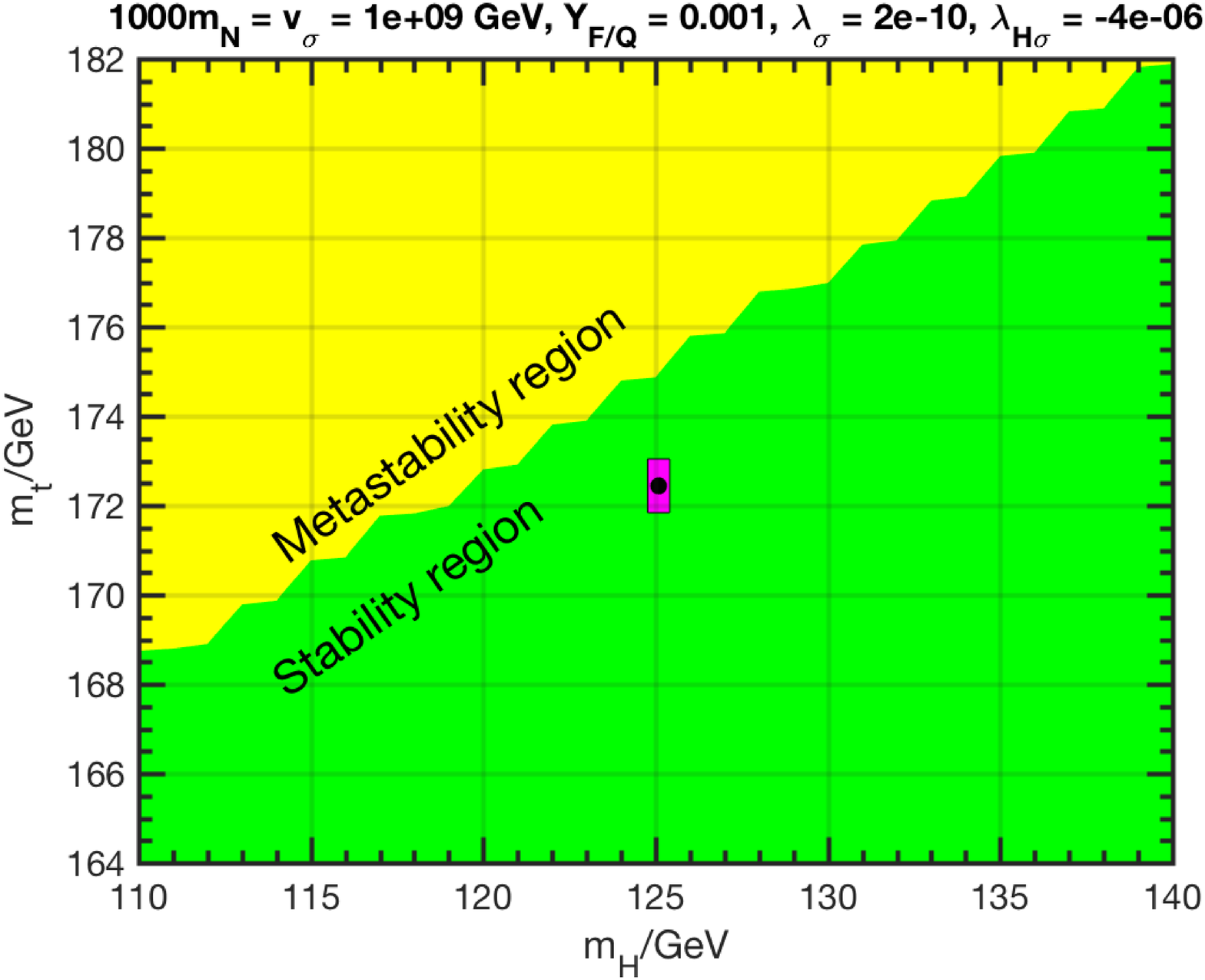}
\includegraphics[width=0.515\textwidth]{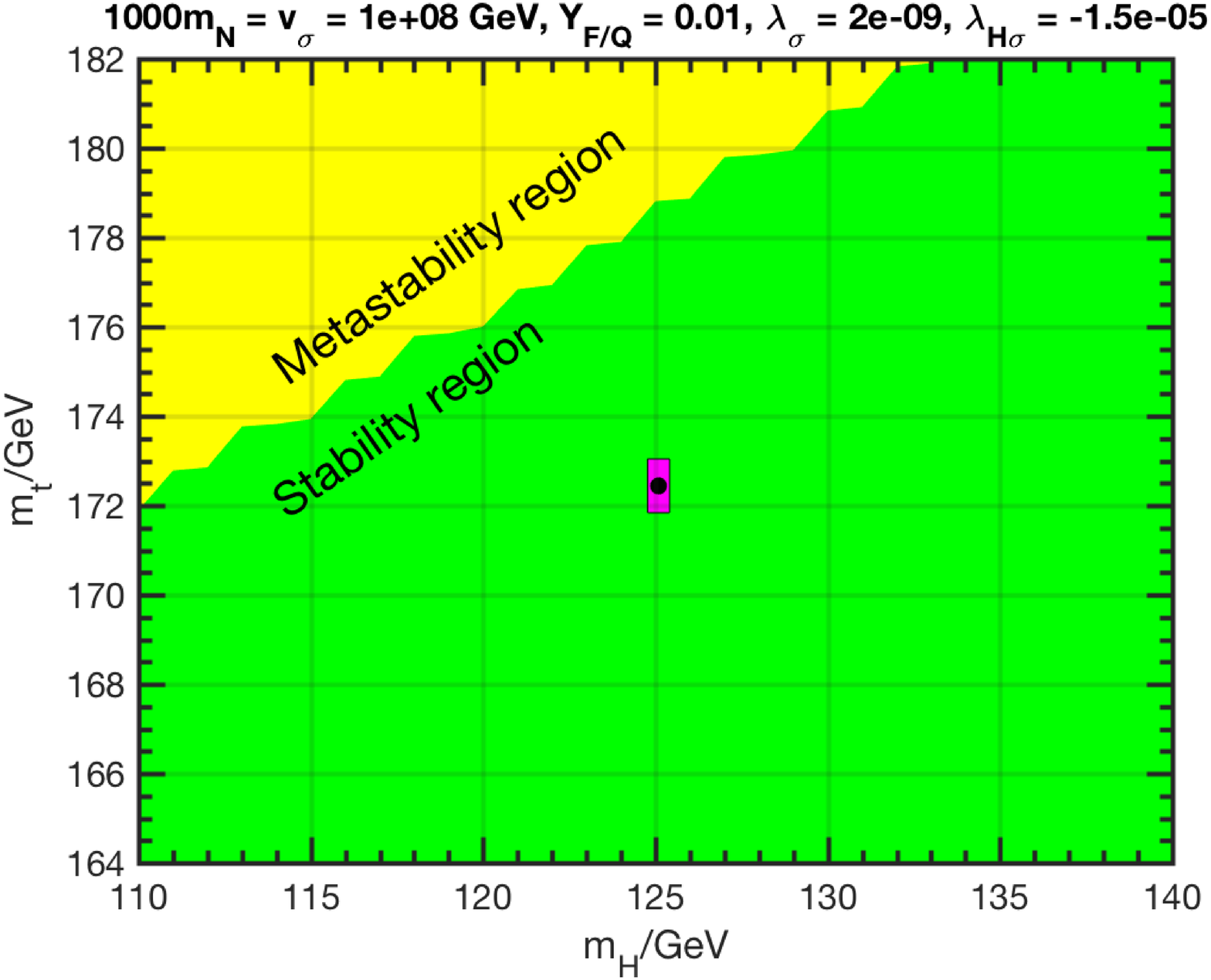}}
\caption{Scalar potential vacuum stability regions for $\lambda_{H\sigma} \approx -10^{-5}$ in ($m_H$, $m_t$) plane.}\label{Fig:3}
\end{figure}

Another way is to solve the SMASH RGE's from the SM scale, not bothering to solve the low-energy SM RGE's at all. We gave both examples in Fig.\ref{Fig:1} and \ref{Fig:2}. In Fig.\ref{Fig:3} we have shown how the current experimental values of $m_t = 172.44 \pm 0.60$ GeV and $m_H = 125.09 \pm 0.32$ GeV are staying within the stability region due to the $\lambda_{H\sigma}\sim -10^{-6}$.

\section{Choice of $\lambda_{H\sigma}$}
To avoid the overproduction of dark radiation via the cosmic axion background, we choose $\lambda_{H\sigma} < 0$. To obtain the observed matter-antimatter asymmetry via leptogenesis, a hierarchy on heavy Majorana neutrinos $N_i$ is required. This is achieved by assigning $Y_n = y_N \times diag(1, 2, 2.1)$, where $Y_N$ and $y_N$ are right-handed and left-handed neutrino Yukawa matrices respectively.

\section{Conclusions}
\begin{enumerate}
\item SMASH unifies axions, seesaw and extended Higgs sector on one energy scale, $\mu\sim 10^{10}$ -- $10^{11}$ GeV, solving several problems badgering the Standard Model in one go.
\item SM vacuum is metastable, since $\lambda_H$ turns negative around $\mu\simeq 10^{12}$ GeV, SMASH can fix this vacuum metastability problem with $\lambda_{H\sigma} \gtrsim -10^{-5}$ at two-loop RGE level.
\item Also, SMASH shows atmospheric neutrino mass splitting is around 0.05 eV and solar neutrino mass splitting is around 0.009 eV.
\end{enumerate}

\end{document}